\documentclass{article}

\usepackage{microtype}
\usepackage{graphicx}
\usepackage{subcaption}
\usepackage{booktabs} 
\usepackage{amsmath}
\usepackage{amssymb}
\usepackage[numbers]{natbib}

\usepackage{hyperref}

\usepackage[accepted]{icml2019}

\newcommand{\ph}{\ \cdot \ }
\newcommand{\pb}{;\ }
\begin{document}

\twocolumn[
\icmltitle{MelNet: A Generative Model for Audio in the Frequency Domain}




\begin{icmlauthorlist}
\icmlauthor{Sean Vasquez}{fb}
\icmlauthor{Mike Lewis}{fb}
\end{icmlauthorlist}

\icmlaffiliation{fb}{Facebook AI Research}

\icmlkeywords{}

\vskip 0.29in
]



\printAffiliationsAndNotice{}  

\begin{abstract}
Capturing high-level structure in audio waveforms is challenging because a single second of audio spans tens of thousands of timesteps.
While long-range dependencies are difficult to model directly in the time domain, we show that they can be more tractably modelled in two-dimensional time-frequency representations such as spectrograms.
By leveraging this representational advantage, in conjunction with a highly expressive probabilistic model and a multiscale generation procedure, we design a model capable of generating high-fidelity audio samples which capture structure at timescales that time-domain models have yet to achieve.
We apply our model to a variety of audio generation tasks, including unconditional speech generation, music generation, and text-to-speech synthesis---showing improvements over previous approaches in both density estimates and human judgments.\looseness=-1
\end{abstract}

\begin{figure*}[t]
    \begin{subfigure}{.49\linewidth}
        \includegraphics[width=\linewidth]{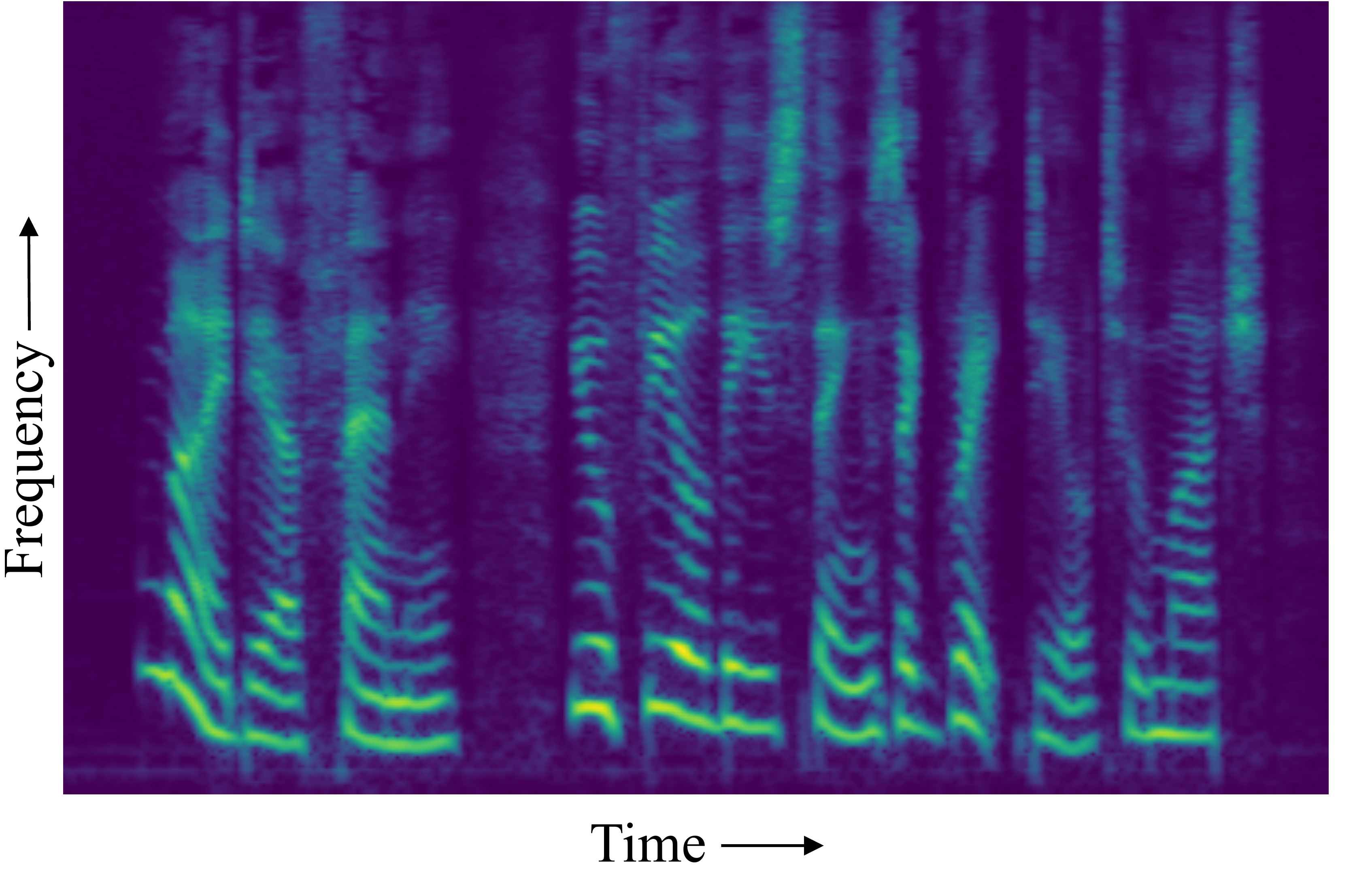}
        \caption{Spectrogram representation}
    \end{subfigure}
    \hfill
    \begin{subfigure}{.49\linewidth}
        \includegraphics[width=\linewidth]{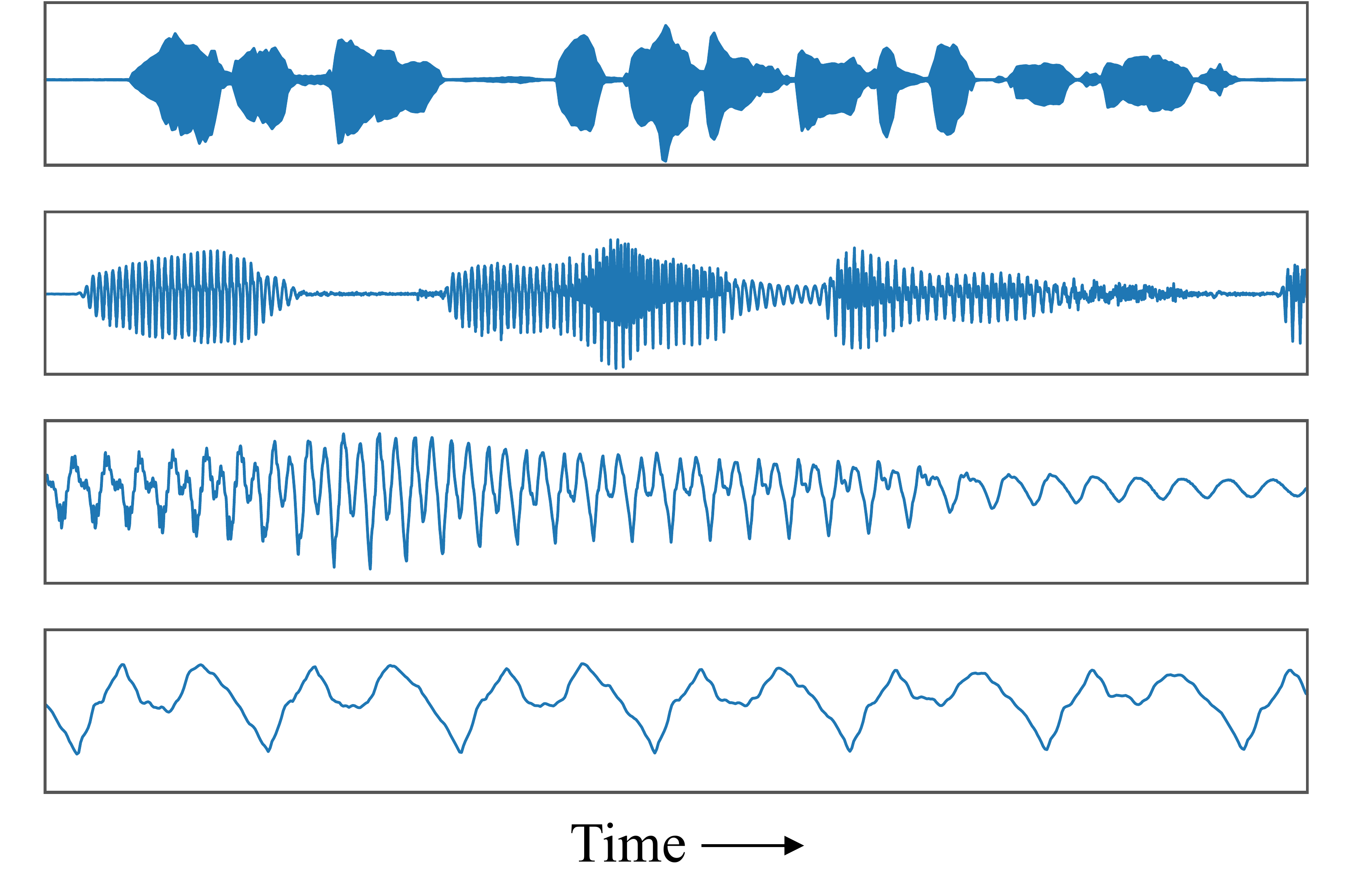}
        \caption{Waveform representation (1x, 5x, 25x, 125x magnifications)}
    \end{subfigure}
    \caption{\label{fig:representation}Spectrogram and waveform representations of the same four-second audio signal. The waveform spans nearly 100,000 timesteps whereas the temporal axis of the spectrogram spans roughly 400.  Complex structure is nested within the temporal axis of the waveform at various timescales, whereas the spectrogram has structure which is smoothly spread across the time-frequency plane.}
\end{figure*}

\section{Introduction}
Audio waveforms have complex structure at drastically varying timescales, which presents a challenge for generative models.  Local structure must be captured to produce high-fidelity audio, while long-range dependencies spanning tens of thousands of timesteps must be captured to generate audio which is globally consistent.  Existing generative models of waveforms such as WaveNet~\citep{oord2016wavenet} and SampleRNN~\citep{mehri2016samplernn} are well-adapted to model local dependencies, but as these models typically only backpropagate through a fraction of a second, they are unable to capture high-level structure that emerges on the scale of several seconds.

We introduce a generative model for audio which captures longer-range dependencies than existing end-to-end models.  We primarily achieve this by modelling 2D time-frequency representations such as spectrograms rather than 1D time-domain waveforms (Figure~\ref{fig:representation}).  The temporal axis of a spectrogram is orders of magnitude more compact than that of a waveform, meaning dependencies that span tens of thousands of timesteps in waveforms only span hundreds of timesteps in spectrograms. In practice, this enables our spectrogram models to generate unconditional speech and music samples with consistency over multiple seconds whereas time-domain models must be conditioned on intermediate features to capture structure at similar timescales.  Additionally, it enables fully end-to-end text-to-speech---a task which has yet to be proven feasible with time-domain models.\looseness=-1

Modelling spectrograms can simplify the task of capturing global structure, but can weaken a model's ability to capture  local characteristics that correlate with audio fidelity.
Producing high-fidelity audio has been challenging for existing spectrogram models, which we attribute to the lossy nature of spectrograms and oversmoothing artifacts which result from insufficiently expressive models.  To reduce information loss, we model high-resolution spectrograms which have the same dimensionality as their corresponding time-domain signals.  To limit oversmoothing, we use a highly expressive autoregressive model which factorizes the distribution over both the time and frequency dimensions.

Modelling both fine-grained details and high-level structure in high-dimensional distributions is known to be challenging for autoregressive models. To capture both local and global structure in spectrograms with hundreds of thousands of dimensions, we employ a multiscale approach which generates spectrograms in a coarse-to-fine manner.  A low-resolution, subsampled spectrogram that captures high-level structure is generated initially, followed by an iterative upsampling procedure that adds high-resolution details.

Combining these representational and modelling techniques yields a highly expressive, broadly applicable, and fully end-to-end generative model of audio.  Our contributions are:\looseness=-1
\begin{itemize}
    \item We introduce MelNet, a generative model for spectrograms which couples a fine-grained autoregressive model and a multiscale generation procedure to jointly capture local and global structure.
    \item We show that MelNet is able to model longer-range dependencies than existing time-domain models.
    \item We demonstrate that MelNet is broadly applicable to a variety of audio generation tasks---capable of unconditional speech generation, music generation, and text-to-speech synthesis, entirely end-to-end.
\end{itemize}

\section{Preliminaries}
We briefly present background regarding spectral representations of audio.  Audio is represented digitally as a one-dimensional, discrete-time signal $y = (y_1, \ldots, y_n)$.   Existing generative models for audio have predominantly focused on modelling these time-domain signals directly.  We instead model spectrograms, which are two-dimensional time-frequency representations which contain information about how the frequency content of an audio signal varies through time.  Spectrograms are computed by taking the squared magnitude of the short-time Fourier transform (STFT) of a time-domain signal, i.e.\ $x = \| \text{STFT}(y) \|^2$.  The value of $x_{ij}$ (referred to as amplitude or energy) corresponds to the squared magnitude of the $j{\text{th}}$ element of the frequency response at timestep $i$.  Each slice $x_{i, *}$ is referred to as a \textit{frame}.  We assume a time-major ordering, but following convention, all figures are displayed transposed and with the frequency axis inverted.

Time-frequency representations such as spectrograms highlight how the tones and pitches within an audio signal vary through time.
Such representations are closely aligned with how humans perceive audio.
To further align these representations with human perception, we convert the frequency axis to the Mel scale and apply an elementwise logarithmic rescaling of the amplitudes.
Roughly speaking, the Mel transformation aligns the frequency axis with human perception of pitch and the logarithmic rescaling aligns the amplitude axis with human perception of loudness.

Spectrograms are lossy representations of their corresponding time-domain signals.
The Mel transformation discards frequency information and the removal of the STFT phase discards temporal information.
When recovering a time-domain signal from a spectrogram, this information loss manifests as distortion in the recovered signal.
To minimize these artifacts and improve the fidelity of generated audio, we model high-resolution spectrograms.
The temporal resolution of a spectrogram can be increased by decreasing the STFT hop size, and the frequency resolution can be increased by increasing the number of mel channels.
Generated spectrograms are converted back to time-domain signals using classical spectrogram inversion algorithms.  We experiment with both Griffin-Lim~\citep{griffin1984signal} and a gradient-based inversion algorithm~\citep{decorsiere2015inversion}, and ultimately use the latter as it generally produced audio with fewer artifacts.

\begin{figure*}[t]\
    \begin{subfigure}[t]{.72\linewidth}
        \centering
        \includegraphics[width=\linewidth]{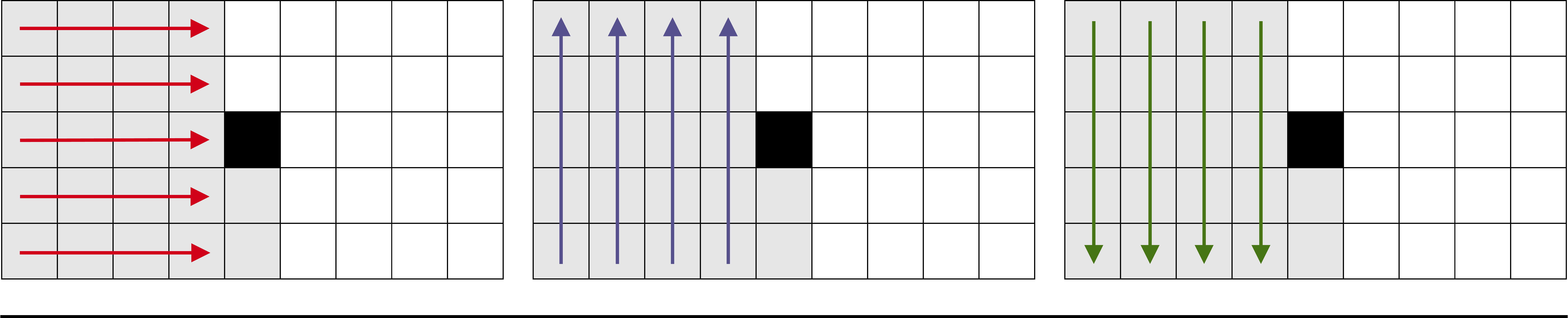}
        \caption{Time-delayed stack}
    \end{subfigure}
    \hfill
    \begin{subfigure}[t]{.232\linewidth}
        \includegraphics[width=\linewidth]{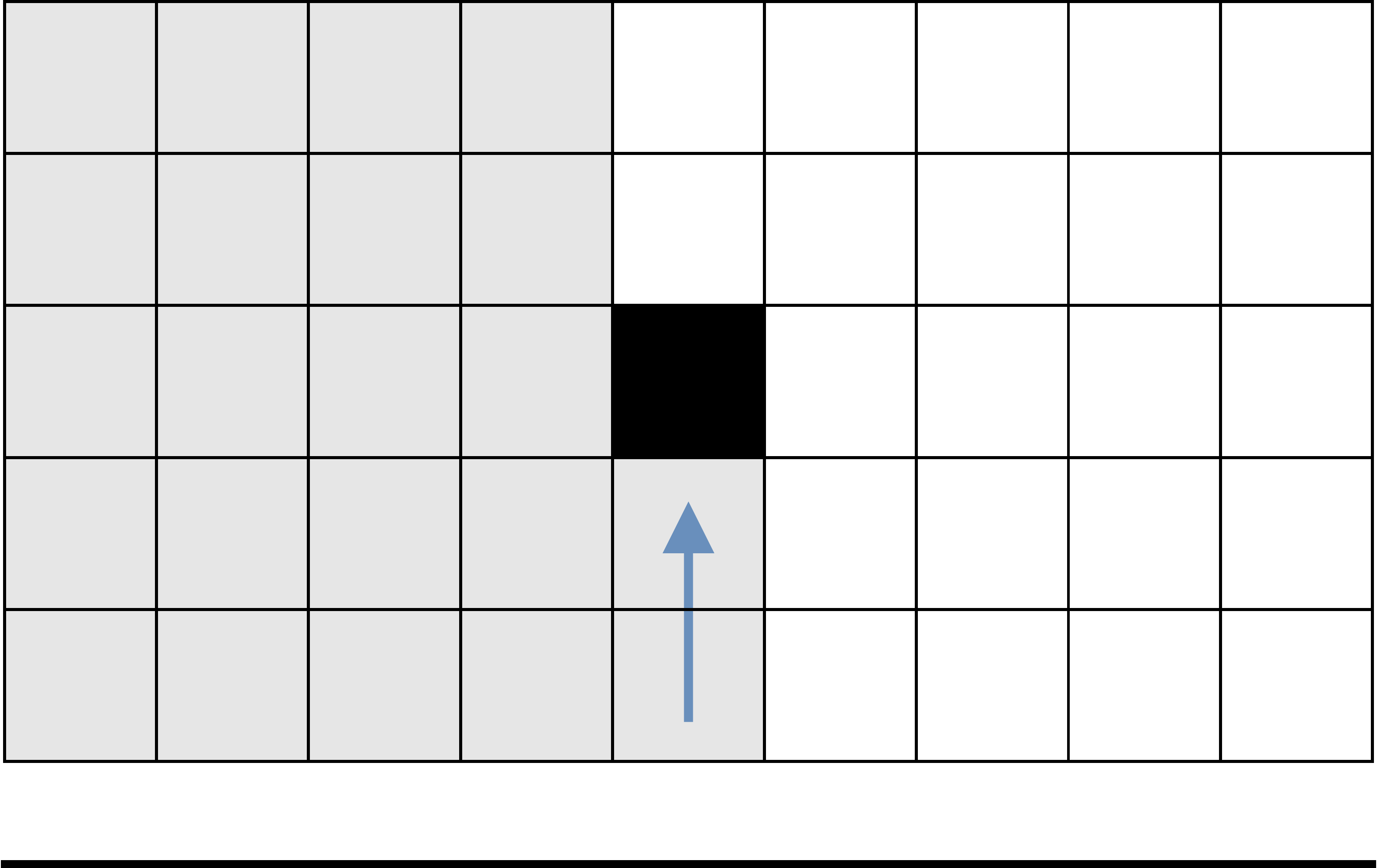}
        \caption{Frequency-delayed stack}
    \end{subfigure}
    \caption{The context $x_{<ij}$ (grey) for the element $x_{ij}$ (black) is encoded using 4 RNNs.  Three of these are used in the time-delayed stack to extract features from all preceding frames.  The fourth is used in the frequency-delayed stack to extract features from all preceding elements within the current frame.  Each arrow denotes an individual RNN cell and arrows of the same color use shared parameters.}
    \label{fig:rnn}
\end{figure*}

\section{Probabilistic Model} \label{sec:model}
We use an autoregressive model which factorizes the joint distribution over a spectrogram $x$ as a product of conditional distributions.
Given an ordering of the dimensions of $x$, we define the context $x_{<ij}$ as the elements of $x$ that precede $x_{ij}$.
We default to a row-major ordering which proceeds through each frame $x_{i, *}$ from low to high frequency, before progressing to the next frame. The joint density is factorized as\looseness=-1
\begin{equation}
p(x) = \prod_{i}\prod_{j}p(x_{ij} \mid x_{<ij} \pb \theta_{ij}),
\end{equation}
where $\theta_{ij}$ parameterizes a univariate density over $x_{ij}$.  We model each factor distribution as a Gaussian mixture model with $K$ components. Thus, $\theta_{ij}$ consists of $3K$ parameters corresponding to means $\{\mu_{ijk}\}_{k=1}^{K}$, standard deviations $\{\sigma_{ijk}\}_{k=1}^{K}$, and mixture coefficients $\{\pi_{ijk}\}_{k=1}^{K}$.  The resulting factor distribution can then be expressed as
\begin{equation}
p(x_{ij} \mid x_{<ij} \pb  \theta_{ij}) = \sum_{k=1}^{K} \pi_{ijk} \ \mathcal{N}(x_{ij} \pb \mu_{ijk}, \sigma_{ijk}).
\end{equation}
Following the work on Mixture Density Networks~\citep{bishop1994mixture} and their application to autoregressive models~\citep{graves2013generating}, $\theta_{ij}$ is modelled as the output of a neural network and computed as a function of the context $x_{<ij}$.  Precisely, for some network $f$ with parameters $\psi$, we have $\theta_{ij} = f(x_{<ij} \pb \psi)$.  A maximum-likelihood estimate for the network parameters is computed by minimizing the negative log-likelihood via gradient descent.

To ensure that the network output parameterizes a valid Gaussian mixture model, the network first computes unconstrained parameters $\{\hat{\mu}_{ijk}, \hat{\sigma}_{ijk}, \hat{\pi}_{ijk}\}_{k=1}^{K}$ as a vector $\hat{\theta}_{ij} \in \mathbb{R}^{3K}$, and enforces constraints on $\theta_{ij}$ by applying the following transformations:
\begin{align}
    \mu_{ijk} &= \hat{\mu}_{ijk} \\
    \sigma_{ijk} &= \exp(\hat{\sigma}_{ijk}) \\
    \pi_{ijk} &= \frac{\exp(\hat{\pi}_{ijk})}{\sum_{k=1}^{K}\exp(\hat{\pi}_{ijk})}\ .
\end{align}
These transformations ensure the standard deviations $\sigma_{ijk}$ are positive and the mixture coefficients $\pi_{ijk}$ sum to one.

\section{Network Architecture}
To model the distribution in an autoregressive manner, we design a network which computes the distribution over $x_{ij}$ as a function of the context $x_{<ij}$.  The network architecture draws inspiration from existing autoregressive models for images~\citep{theis2015generative,oord2016pixel,van2016conditional,chen2017pixelsnail,salimans2017pixelcnn++,parmar2018image,child2019generating}.  In the same way that these models estimate a distribution pixel-by-pixel over the spatial dimensions of an image, our model estimates a distribution element-by-element over the time and frequency dimensions of a spectrogram.  A noteworthy distinction is that spectrograms are not invariant to translation along the frequency axis, making the use of 2D convolution undesirable.  Utilizing multidimensional recurrence instead of 2D convolution has been shown to be beneficial when modelling spectrograms in discriminative settings~\citep{li2016exploring,sainath2016modeling}, which motivates our use of an entirely recurrent architecture.\looseness=-1

Similar to Gated PixelCNN~\citep{van2016conditional}, the network has multiple \textit{stacks} of computation.  These stacks extract features from different segments of the input to collectively summarize the full context $x_{<ij}$:
\begin{itemize}
\item The \textit{time-delayed} stack computes features which aggregate information from all previous frames $x_{<i, *}$.
\item The \textit{frequency-delayed} stack utilizes all preceding elements within a frame, $x_{i, <j}$, as well as the outputs of the time-delayed stack, to compute features which summarize the full context $x_{<ij}$.
\end{itemize}
The stacks are connected at each layer of the network, meaning that the features generated by layer $l$ of the time-delayed stack are used as input to layer $l$ of the frequency-delayed stack.  To facilitate the training of deeper networks, both stacks use residual connections~\citep{he2016deep}.
The outputs of the final layer of the frequency-delayed stack are used to compute the unconstrained parameters $\hat{\theta}$.

\subsection{Time-Delayed Stack}
The time-delayed stack utilizes multiple layers of multi\-dimensional RNNs to extract features from $x_{<i, *}$, the two-dimensional region consisting of all frames preceding $x_{ij}$.  Each multidimensional RNN is composed of three one-dimensional RNNs: one which runs forwards along the frequency axis, one which runs backwards along the frequency axis, and one which runs forwards along the time axis.  Each RNN runs along each slice of a given axis, as shown in Figure~\ref{fig:rnn}.  The output of each layer of the time-delayed stack is the concatenation of the three RNN hidden states.\looseness=-1

We denote the function computed at layer $l$ of the time-delayed stack (three RNNs followed by concatenation) as $\mathcal{F}^{t}_{l}$.  At each layer, the time-delayed stack uses the feature map from the previous layer, $h^{t}[l-1]$, to compute the subsequent feature map $\mathcal{F}^{t}_{l}\big(h^{t}[l - 1]\big)$ which consists of the three concatenated RNN hidden states.  When using residual connections, the computation of $h^{t}[l]$ from $h^{t}[l - 1]$ becomes\looseness=-1
\begin{equation}
h^{t}_{ij}[l] = W^{t}_{l}\mathcal{F}^{t}_{l}\big(h^{t}[l - 1]\big)_{ij} + h^{t}_{ij}[l-1] .
\end{equation}
To ensure the output $h^{t}_{ij}[l]$ is only a function of frames which lie in the context $x_{<ij}$, the inputs to the time-delayed stack are shifted backwards one step in time:
\begin{equation}
h^{t}_{ij}[0] = W^{t}_{0}x_{i-1, j}. \label{eq:init1}
\end{equation}

\subsection{Frequency-Delayed Stack}
The frequency-delayed stack is a one-dimensional RNN which runs forward along the frequency axis.  Much like existing one-dimensional autoregressive models (language models, waveform models, etc.), the frequency-delayed stack operates on a one-dimensional sequence (a single frame) and estimates the distribution for each element conditioned on all preceding elements.  The primary difference is that it is also conditioned on the outputs of the time-delayed stack, allowing it to use the full two-dimensional context $x_{<ij}$.

We denote the function computed by the frequency-delayed stack as $\mathcal{F}^{f}_{l}$.  At each layer, the frequency-delayed stack takes two inputs: the the previous-layer outputs of the frequency-delayed stack, $h^{f}[l - 1]$, and the current-layer outputs of the time-delayed stack $h^{t}[l]$.  These inputs are summed and used as input to a one-dimensional RNN to produce the output feature map $\mathcal{F}^{f}_{l}\big(h^{f}[l - 1], \ h^{t}[l]\big)$ which consists of the RNN hidden state:
\begin{align}
h^{f}_{ij}[l] &= W^{f}_{l}\mathcal{F}^{f}_{l}\big(h^{f}[l - 1], \ h^{t}[l]\big)_{ij} + h^{f}_{ij}[l-1] .
\end{align}
To ensure that $h^{f}_{ij}[l]$ is computed using only elements in the context $x_{<ij}$, the inputs to the frequency-delayed stack are shifted backwards one step along the frequency axis:
\begin{align}
h^{f}_{ij}[0] &= W^{f}_{0}x_{i, j-1}. \label{eq:init2}
\end{align}
At the final layer, layer $L$, a linear map is applied to the output of the frequency-delayed stack to produce the unconstrained parameters:
\begin{equation}
\hat{\theta}_{ij}=W_{\theta}h^{f}_{ij}[L] .
\end{equation}

\subsection{Centralized Stack}
The recurrent state of the time-delayed stack is distributed across an array of RNN cells which tile the frequency axis.  To allow for a more centralized representation, we optionally include an additional stack consisting of an RNN which, at each timestep, takes an entire frame as input and outputs a single vector consisting of the RNN hidden state.
Denoting this function as $\mathcal{F}^{c}_{l}$ gives the layer update
\begin{equation}
h^{c}_{i}[l] = W^{c}_{l}\mathcal{F}^{c}_{l}\big(h^{c}[l - 1]\big)_{i} + h^{c}_{i}[l-1].
\end{equation}
Similar to the time-delayed stack, the centralized stack operates on frames which are shifted backwards one step along the time axis:
\begin{equation}
h^{c}_{i}[0] = W^{c}_{0}x_{i-1, *}.
\end{equation}
The output of the centralized stack is input to the frequency-delayed stack at each layer, meaning that the frequency-delayed stack is a function of three inputs: $h^{f}[l-1]$, $h^{t}[l]$, and $h^{c}[l]$. These three inputs are simply summed and used as input to the RNN in the frequency-delayed stack.

\subsection{Conditioning} \label{sec:cond}
To incorporate conditioning information into the model, conditioning features $z$ are simply projected onto the input layer along with the inputs $x$, altering Equations~\ref{eq:init1} and~\ref{eq:init2}:
\begin{align}
h^{t}_{ij}[0] &= W^{t}_{0}x_{i-1, j} + W^{t}_{z}z_{ij} \\
h^{f}_{ij}[0] &= W^{f}_{0}x_{i, j-1} + W^{f}_{z}z_{ij}.
\end{align}
Reshaping, upsampling, and broadcasting can be used as necessary to ensure the conditioning features have the same time and frequency shape as the input spectrogram, e.g. a one-hot vector representation for speaker ID would first be broadcast along both the time and frequency axes.

\begin{figure}[t]
    \centering
    \includegraphics[width=.88\linewidth]{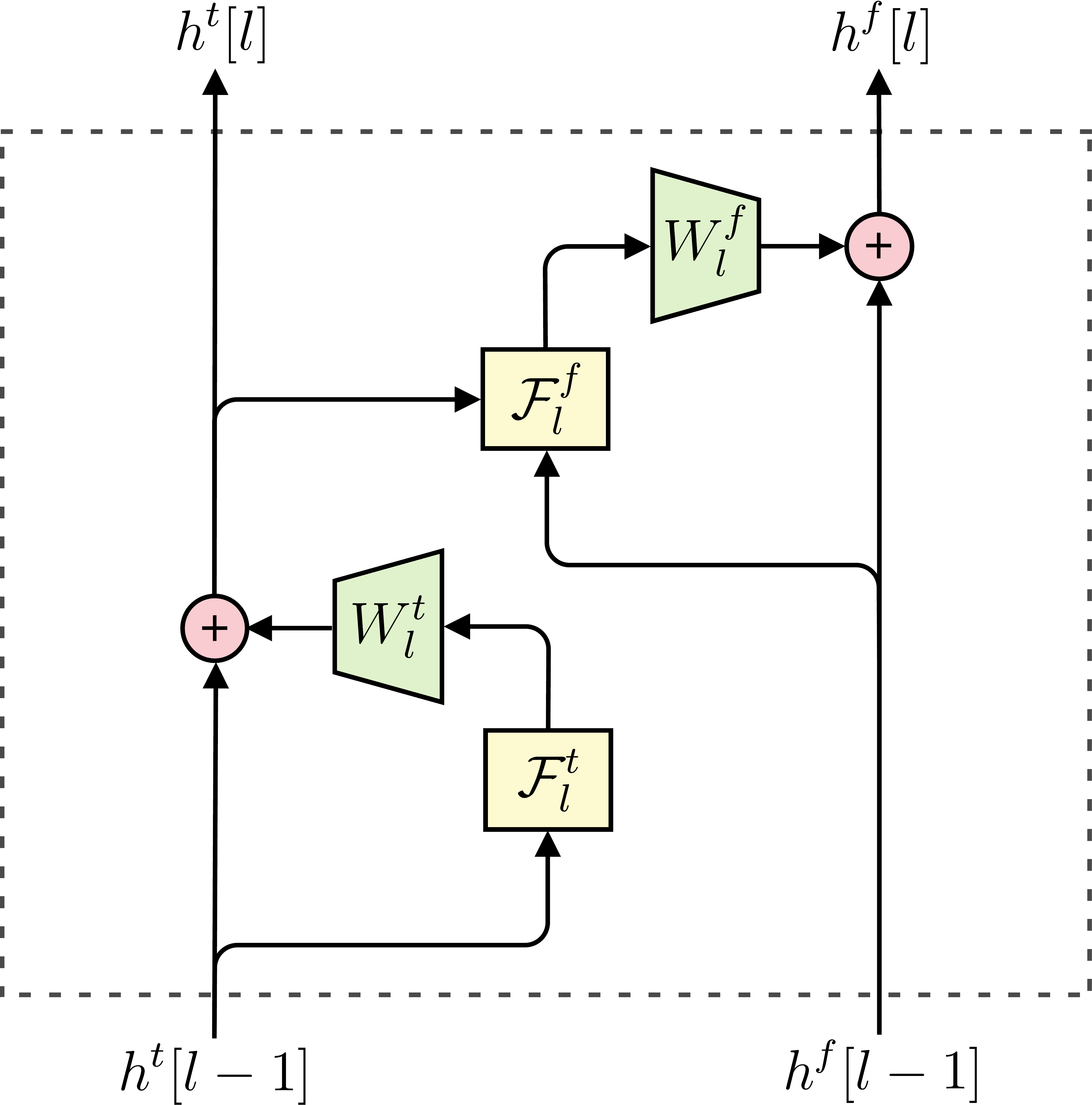}
    \caption{Computation graph for a single layer of the network.  $\mathcal{F}^{t}_{l}$ and $\mathcal{F}^{f}_{l}$ are the functions computed by the time-delayed stack and frequency-delayed stack, respectively, at layer $l$. The outputs of these functions are projected (by the matrices $W^{t}_{l}$ and $W^{f}_{l}$) and summed with the layer inputs to form residual blocks.}
\end{figure}

\section{Learned Alignment}\label{sec:alignment}
For the task of end-to-end text-to-speech, the network must learn a latent alignment between spectrogram frames $(x_{1, *}, \ldots, x_{T, *})$ and discrete character tokens $(c_1, \ldots, c_U)$.  To facilitate this, we first extract character features $(\tilde{c}_1, \ldots, \tilde{c}_U)$ by embedding each character $c_u$ and running a bidirectional RNN over the embeddings.  Extracting character features eases the alignment process by allowing the network to learn both phonetic features which are important for pronunciation and higher-level semantics which must be understood to infer proper intonation and prosody.

We use an attention mechanism which is a straightforward variant of the location-based Gaussian mixture attention introduced by \citet{graves2013generating}.  The attention mechanism consists of an RNN in the centralized stack which, at timestep $i$, computes an attention vector $w_i$ as a weighted sum of character features $(\tilde{c}_1,  \ldots, \tilde{c}_U)$.  The weights correspond to a learned attention distribution $\phi_i(\ph \pb \gamma_i)$ whose parameters $\gamma_i$ are computed as a simple function $g$ of the RNN hidden state.  This is expressed by the following recurrence, where $y_i$ represents an arbitrary input at timestep $i$:
\begin{align}
    h_i &= \text{RNN}\Big([y_i, \  w_{i - 1}], \  h_{i-1}\Big) \\
    w_i &= \sum_{u=1}^{U}\phi_i\Big(u \pb \gamma_i= g\big(h_{i}\big)\Big)\tilde{c}_u .
\end{align}
The original formulation parameterizes $\phi_i(\ph \pb \gamma_i)$ as an unnormalized Gaussian mixture model, whereas we use a discretized mixture of logistics~\citep{salimans2017pixelcnn++}.  In either case, the distribution is parameterized by $\gamma_i=\{\kappa_i^m, \beta_i^m, \alpha_i^m\}_{m=1}^M$, corresponding to $M$ means, scales, and mixture coefficients. We define the function $g$ as a trainable linear mapping of the RNN hidden state $h_{i}$ followed by transformations which constrain the mixture coefficients $\alpha_i^m$ to sum to one, the scales $\beta_i^m$ to be positive, and the means $\kappa_i^m$ to be monotonically increasing with $i$:
\begin{align}
    \{\hat{\kappa}_i^m, \hat{\beta}_i^m, \hat{\alpha}_i^m\}_{m=1}^M &= W_{g}h_{i}\\
    \kappa_i^m &= \kappa_{i-1}^m + \exp(\hat{\kappa}_i^m) \\
    \beta_i^m &= \exp(\hat{\beta}_i^m) \\
    \alpha_i^m &= \frac{\exp(\hat{\alpha}_i^m)}{\sum_{m=1}^M\exp(\hat{\alpha}_i^m)}.
\end{align}
The resulting mixture of logistics distribution parameterized by $\gamma_i$ has the distribution function
\begin{equation}
F_i(u \pb \gamma_i) = \sum_{m=1}^M \alpha_i^m \bigg(1 + \exp\!\bigg(\frac{\kappa_i^m - u}{\beta_i^m}\bigg)\bigg)^{\!-1}
\end{equation}
which is then used to compute the discretized attention distribution
\begin{equation}
\phi_i(u \pb \ \gamma_i) = F_i(u + 0.5 \pb \gamma_i) - F_i(u - 0.5 \pb \gamma_i).
\end{equation}
The network needs a criterion by which it can determine whether it has finished `reading' the text and can terminate sampling.  If we interpret $\phi_i(u  \pb \gamma_i)$ as the network's belief that it is reading character $c_u$ at timestep $i$, then the network's belief that it has passed the final character $c_U$ is $\sum_{U + 1}^{\infty}\phi_i(u  \pb \gamma_i)$, which can be expressed in closed form as $\overline{F}_i(U + 0.5 \pb \gamma_i)$ where $\overline{F}_i$ is the survival function $1 - F_i$.  We stop sampling based on a simple threshold of this value, terminating at the first timestep $i$ such that $\overline{F}_i(U + 0.5 \pb \gamma_i) > \tau$.  We compute an estimate for $\tau$ after the network is trained, using the empirical mean $\hat{\tau} = \frac{1}{N}\sum_{n=1}^{N}\overline{F}_{T_n}(U_n + 0.5 \pb \gamma_{T_n})$.

\begin{figure}[t]
    \centering
    \includegraphics[width=\linewidth]{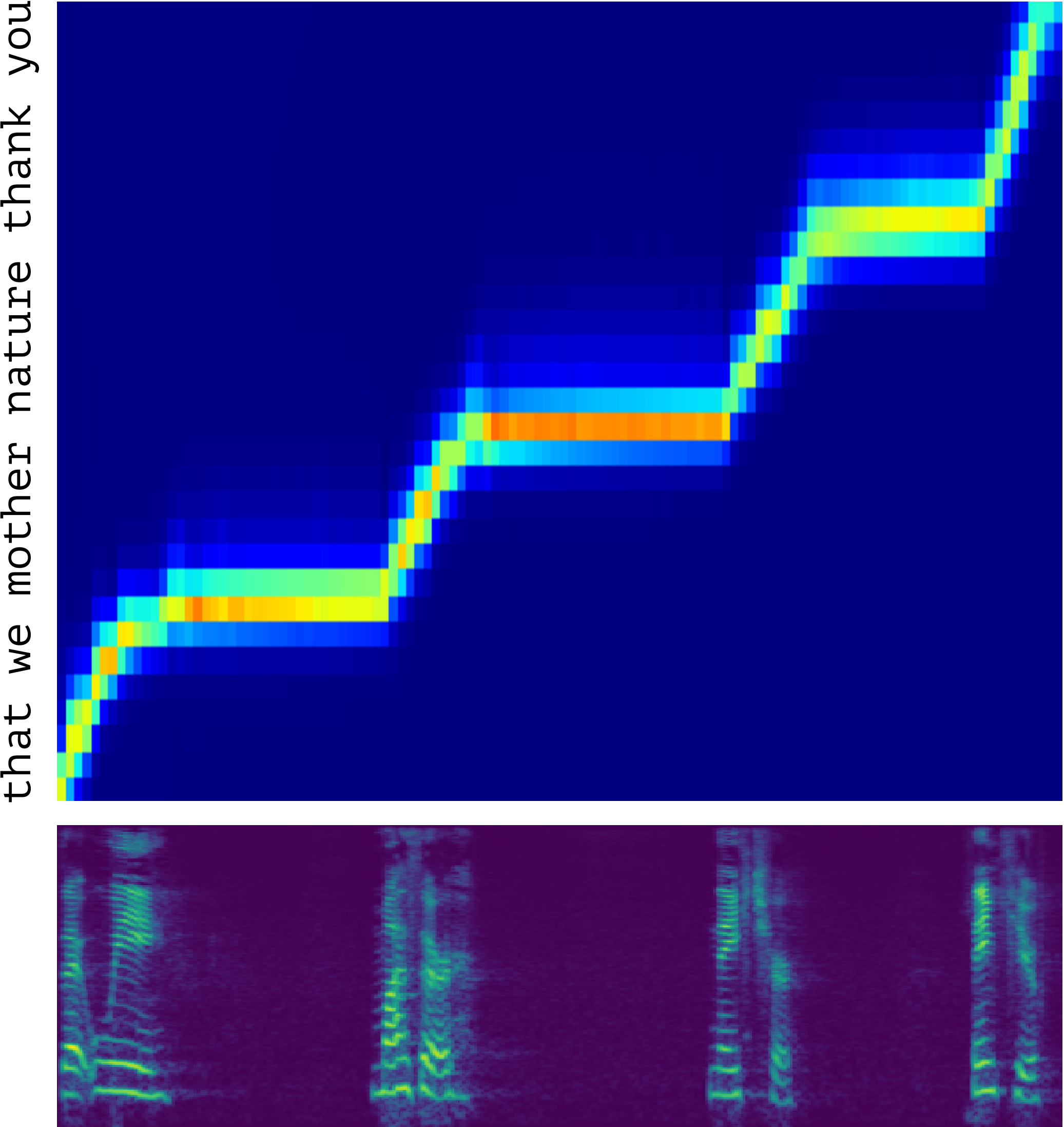}\\
    \caption{Learned alignment between a spectrogram and the character sequence \emph{that we mother nature thank you}.  Column $i$ corresponds to the learned attention distribution $\phi_i(\ph \pb \gamma_i)$.  The text is read with long, deliberate pauses (\emph{that we / mother / nature / thank you}) which appear as flat regions in the alignment.}
\end{figure}

\begin{figure*}[t]
    \begin{subfigure}[t]{.32\linewidth}
        \centering
        \includegraphics[width=\linewidth]{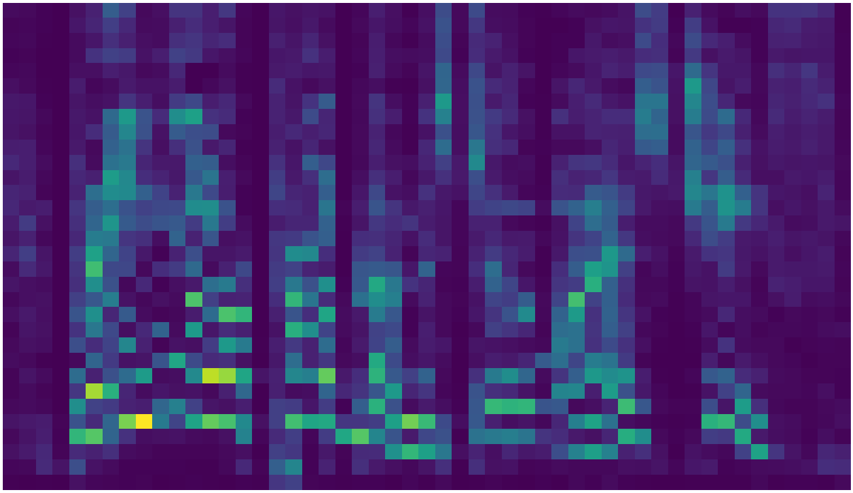}
        \caption{Tier 1 \ ($32 \times 50$)}
    \end{subfigure}
    \hfill
    \begin{subfigure}[t]{.32\linewidth}
        \centering
        \includegraphics[width=\linewidth]{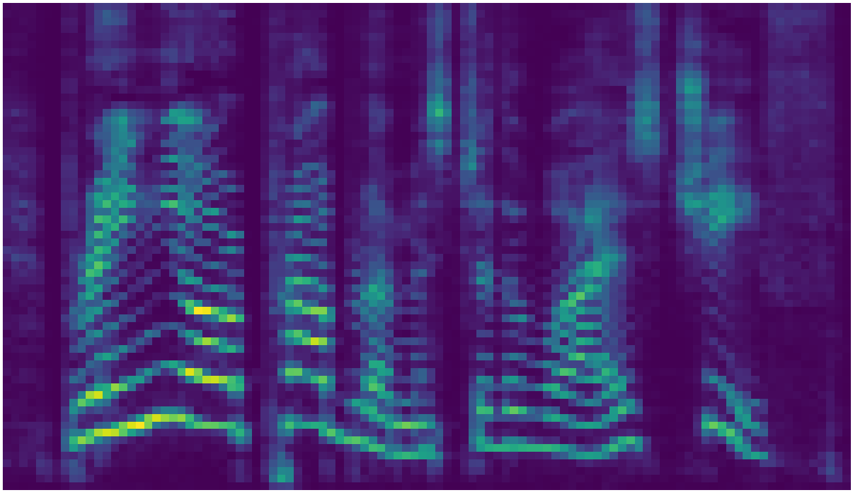}
        \caption{Tiers 1--3 \ ($64 \times 100$)}
    \end{subfigure}
    \hfill
    \begin{subfigure}[t]{.32\linewidth}
        \centering
        \includegraphics[width=\linewidth]{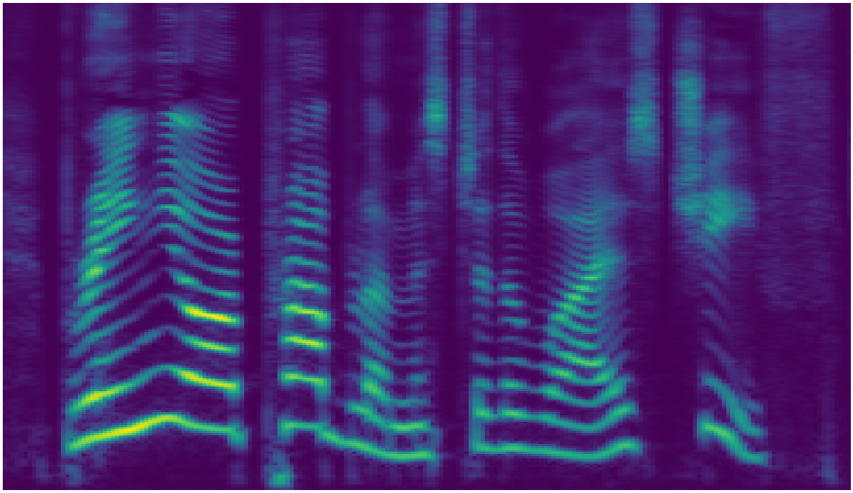}
        \caption{Tiers 1--6 \ ($256 \times 200$)}
    \end{subfigure}
    \caption{A sampled spectrogram viewed at different stages of the multiscale generation procedure. The initial tier dictates high-level structure and subsequent tiers add fine-grained details.  Each upsampling tier doubles the resolution of the spectrogram, resulting in the initial tier being upsampled by a factor of 32.}
    \label{fig:multiscale1}
\end{figure*}

\begin{figure*}[t]
    \centering
    \begin{subfigure}[t]{.11\linewidth}
        \centering
        \includegraphics[width=.87\linewidth]{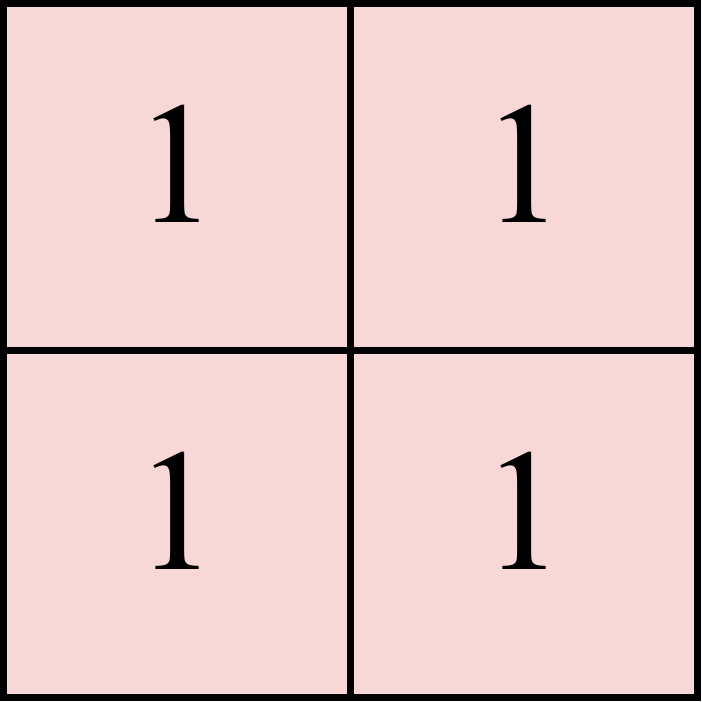}
        \caption{$p(x^1 \pb \psi^1)$}
    \end{subfigure}
    \hfill
    \begin{subfigure}[t]{.23\linewidth}
        \centering
        \includegraphics[width=\linewidth]{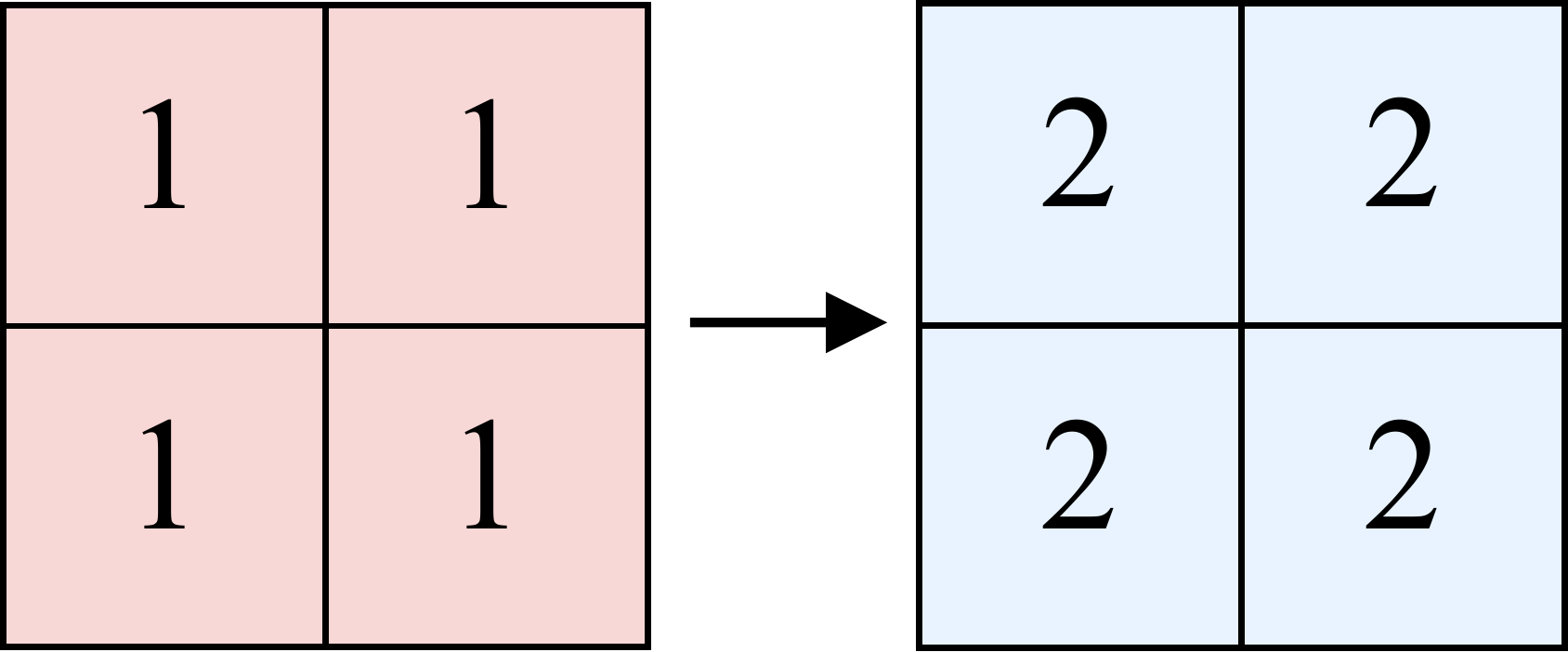}
        \caption{$p(x^2 \mid x^1 \pb \psi^2)$}
    \end{subfigure}
    \hfill
    \begin{subfigure}[t]{.23\linewidth}
        \centering
        \includegraphics[width=\linewidth]{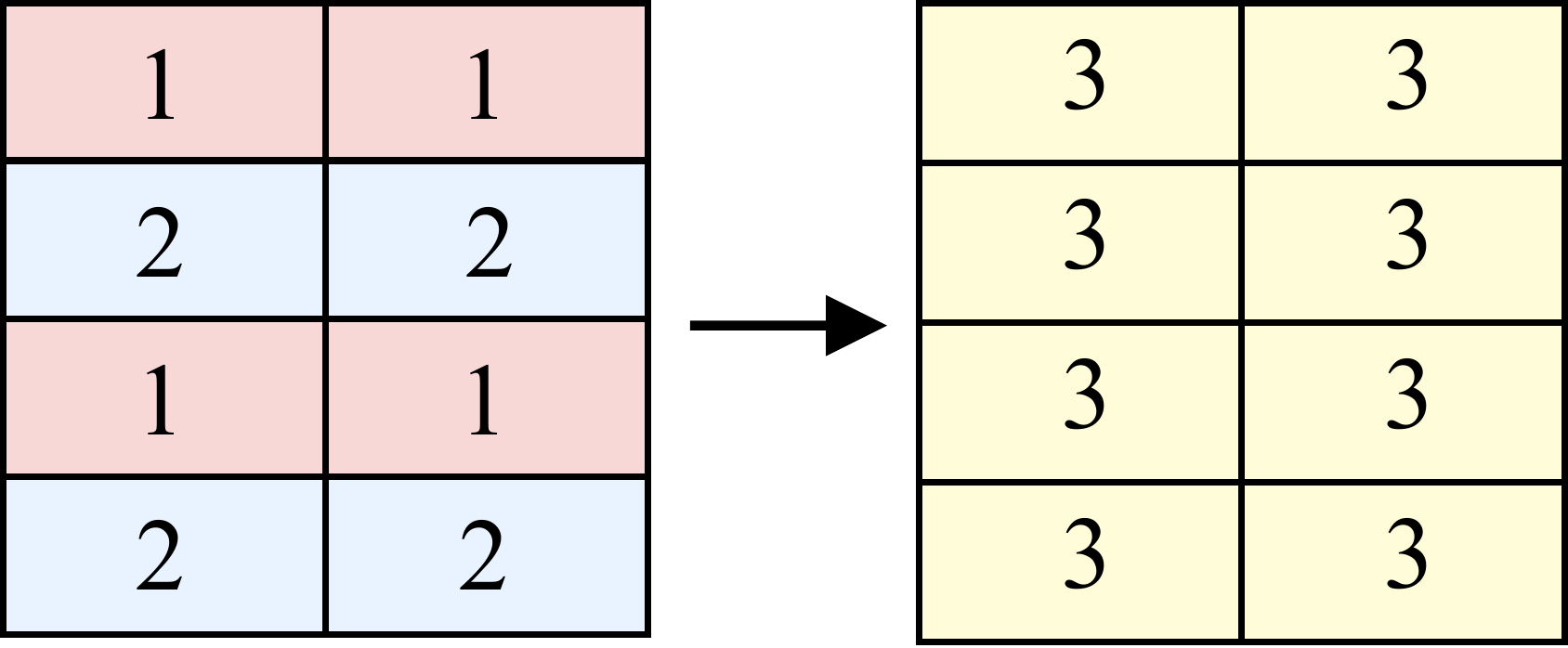}
        \caption{$p(x^3 \mid x^1, x^2 \pb \psi^3)$}
    \end{subfigure}
    \hfill
    \begin{subfigure}[t]{.23\linewidth}
        \centering
        \includegraphics[width=\linewidth]{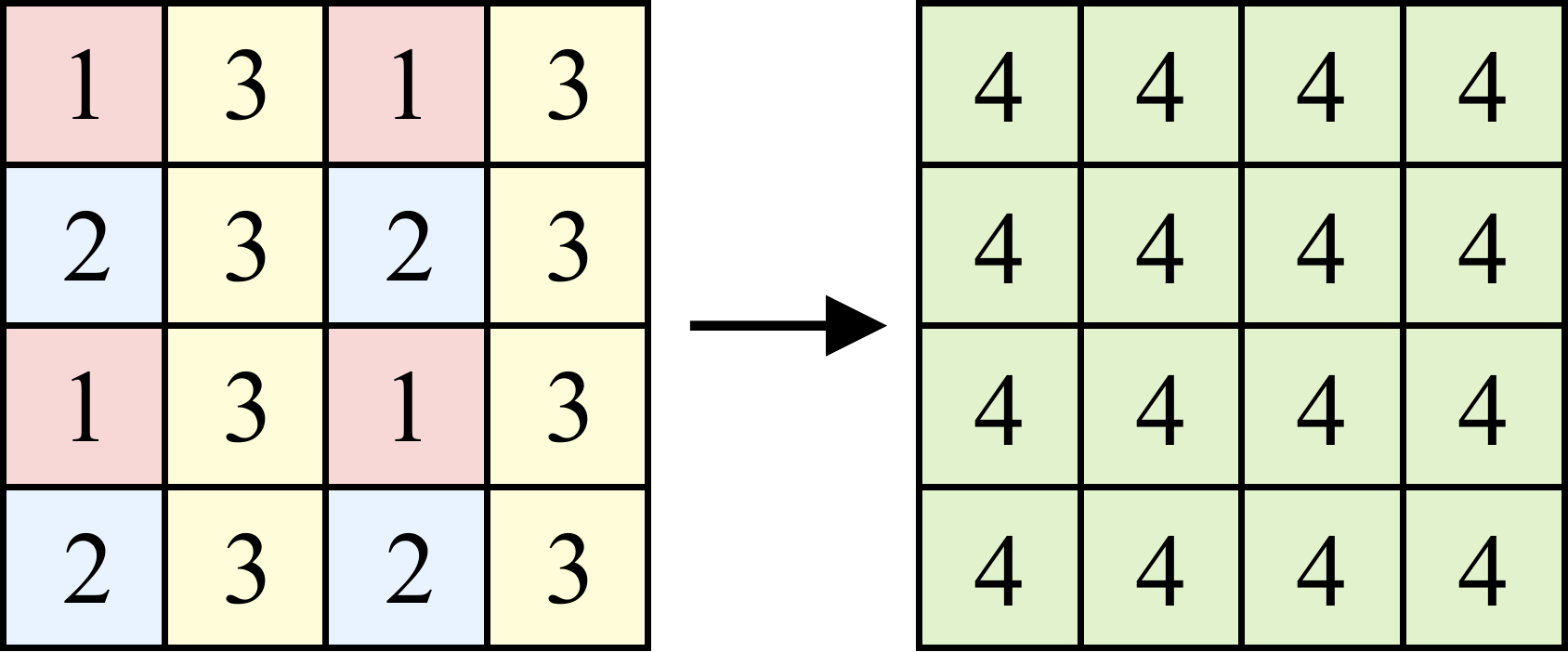}
        \caption{$p(x^4 \mid x^1, x^2, x^3 \pb \psi^4)$}
    \end{subfigure}
    \caption{Schematic showing how tiers of the multiscale model are interleaved and used to condition the distribution for the subsequent tier.  a) The initial tier is generated unconditionally.  b) The second tier is generated conditionally given the the initial tier.  c) The outputs of tiers 1 and 2 are interleaved along the frequency axis and used to condition the generation of tier 3.  d) Tier 3 is interleaved along the time axis with all preceding tiers and used to condition the generation of tier 4.}
    \label{fig:multiscale2}
\end{figure*}

\section{Multiscale Modelling}\label{sec:multiscale}
To improve audio fidelity, we generate high-resolution spectrograms which have the same dimensionality as their corresponding time-domain representations. Under this regime, a single training example has several hundreds of thousands of dimensions.  Capturing global structure in such high-dimensional distributions is challenging for autoregressive models, which are biased towards capturing local dependencies.  To counteract this, we utilize a multiscale approach which effectively permutes the autoregressive ordering so that a spectrogram is generated in a coarse-to-fine order.\looseness=-1

The elements of a spectrogram $x$ are partitioned into $G$ tiers $x^{1}, \ldots, x^{G}$, such that each successive tier contains higher-resolution information.  We define $x^{<g}$ as the union of all tiers which precede $x^g$, i.e. $x^{<g}~=~(x^{1},\ldots, x^{g-1})$.  The distribution is factorized over tiers:
\begin{equation}
p(x \pb \psi) = \prod_{g}p(x^{g} \mid x^{<g} \pb \psi^{g}),
\end{equation}
and the distribution of each tier is further factorized element-by-element as described in Section~\ref{sec:model}.  We explicitly include the parameterization by $\psi = (\psi^{1}, \ldots, \psi^{G})$ to indicate that each tier is modelled by a separate network.

\subsection{Training}
During training, the tiers are generated by recursively partitioning a spectrogram into alternating rows along either the time or frequency axis.  We define a function \texttt{split} which partitions an input into even and odd rows along a given axis.  The initial step of the recursion applies the \texttt{split} function to a spectrogram $x$, or equivalently $x^{<G+1}$, so that the even-numbered rows are assigned to $x^G$ and the odd-numbered rows are assigned to $x^{<G}$.  Subsequent tiers are defined similarly in a recursive manner:\looseness=-1
\begin{equation}
x^{g}, \ x^{<g} = \texttt{split}(x^{<g+1}).
\end{equation}
At each step of the recursion, we model the distribution $p(x^{g} \mid x^{<g} \pb \psi^{g})$.  The final step of the recursion models the unconditional distribution over the initial tier $p(x^{1} \pb \psi^{1})$.

To model the conditional distribution $p(x^{g} \mid x^{<g} \pb \psi^{g})$, the network at each tier needs a mechanism to incorporate information from the preceding tiers $x^{<g}$.  To this end, we add a feature extraction network which computes features from $x^{<g}$ which are used to condition the generation of $x^{g}$.  We use a multidimensional RNN consisting of four one-dimensional RNNs which run bidirectionally along slices of both axes of the context $x^{<g}$.  A layer of the feature extraction network is similar to a layer of the time-delayed stack, but since the feature extraction network is not causal, we include an RNN which runs backwards along the time axis and do not shift the inputs.  The hidden states of the RNNs in the feature extraction network are used to condition the generation of $x^{g}$.  Since each tier doubles the resolution, the features extracted from $x^{<g}$ have the same time and frequency shape as $x^{g}$, allowing the conditioning mechanism described in section~\ref{sec:cond} to be used straightforwardly.

\subsection{Sampling} To sample from the multiscale model we iteratively sample a value for $x^{g}$ conditioned on $x^{<g}$ using the learned distributions defined by the estimated network parameters $\hat{\psi} = (\hat{\psi}^{1}, \ldots, \hat{\psi}^{G})$.  The initial tier, $x^{1}$, is generated unconditionally by sampling from $p(x^{1} \pb \hat{\psi}^{1})$ and subsequent tiers are sampled from  $p(x^{g} \mid x^{<g} \pb \hat{\psi}^{g})$.  At each tier, the sampled $x^{g}$ is interleaved with the context $x^{<g}$:
\begin{equation}
x^{<g+1} = \texttt{interleave}(x^{g}, \ x^{<g}).
\end{equation}
The \texttt{interleave} function is simply the inverse of the \texttt{split} function.  Sampling terminates once a full spectrogram, $x^{<G+1}$, has been generated.  A spectrogram generated by a multiscale model is shown in Figure~\ref{fig:multiscale1} and the sampling procedure is visualized schematically in Figure~\ref{fig:multiscale2}.

\begin{table*}[t]
\centering
\begin{tabular}{lccccccc}\toprule
                               &\hspace{0pt}   &\multicolumn{3}{c}{Unconditional}             &\hspace{16pt}   &\multicolumn{2}{c}{Text-to-Speech} \\
                                               \cmidrule(lr){3-5}                                            \cmidrule(lr){7-8}
                               &\hspace{0pt}   & Blizzard     & MAESTRO      & VoxCeleb2   &\hspace{16pt}   & Blizzard      & TED-LIUM 3    \\\midrule
Tiers                          &\hspace{0pt}   & 6            & 4            & 5           &\hspace{16pt}   & 6             & 5             \\
Layers (Initial Tier)          &\hspace{0pt}   & 12           & 16           & 16          &\hspace{16pt}   & 8             & 12            \\
Layers (Upsampling Tiers)      &\hspace{0pt}   & 5-4-3-2-2    & 6-5-4        & 6-5-4-3     &\hspace{16pt}   & 5-4-3-2-2     & 6-5-4-3       \\
Hidden Size                    &\hspace{0pt}   & 512          & 512          & 512         &\hspace{16pt}   & 512           & 512           \\
GMM Mixture Components         &\hspace{0pt}   & 10           & 10           & 10          &\hspace{16pt}   & 10            & 10            \\
Attention Mixture Components   &\hspace{0pt}   & -            & -            & -           &\hspace{16pt}   & 10            & 10            \\\midrule
Batch Size                     &\hspace{0pt}   & 32           & 16           & 128         &\hspace{16pt}   & 32            & 64            \\
Sample Rate (Hz)               &\hspace{0pt}   & 22,050       & 22,050       & 16,000      &\hspace{16pt}   & 22,050        & 16,000        \\
Max Sample Duration (s)        &\hspace{0pt}   & 10           & 6            & 6           &\hspace{16pt}   & 10            & 10            \\
Mel Channels                   &\hspace{0pt}   & 256          & 256          & 180         &\hspace{16pt}   & 256           & 180           \\
STFT Hop Size                  &\hspace{0pt}   & 256          & 256          & 180         &\hspace{16pt}   & 256           & 180           \\
STFT Window Size               &\hspace{0pt}   & $6\cdot256$  & $6\cdot256$  &$6\cdot180$  &\hspace{16pt}   & $6\cdot256$   & $6\cdot180$   \\\bottomrule
\end{tabular}
\caption{\label{table:melnet}MelNet hyperparameters.
    All RNNs use LSTM cells~\citep{hochreiter1997long}.
    All models are trained with RMSProp~\citep{tieleman2012lecture} with a learning rate of $10^{-4}$ and momentum of $0.9$.
    The initial values for all recurrent states are trainable parameters.
    A single hyperparameter controls the width of the network---all hidden sizes (RNN state size, residual connections, embeddings, etc.) are defined by a single value, denoted \emph{hidden size} in the table.
    Only the initial tier is conditioned on text and speaker ID.
    Only the initial tier uses a centralized stack.
    When an attention cell is used in the centralized stack, it is inserted in the middle layer $(L/2)$.}
\end{table*}

\section{Experiments}
To demonstrate the MelNet is broadly applicable as a generative model for audio, we train the model on a diverse set of audio generation tasks using four publicly available datasets.  We explore three unconditional audio generation tasks (single-speaker speech generation, multi-speaker speech generation, and music generation) as well as two text-to-speech tasks (single-speaker TTS and multi-speaker TTS).  Generated audio samples for each task are available on the accompanying web page.\hspace{-.1em}\footnote{\url{https://audio-samples.github.io}}  We include samples generated using the priming and biasing procedures described by~\citet{graves2013generating}. Biasing lowers the temperature of the distribution at each timestep and priming seeds the model state with a given sequence of audio prior to sampling.

\subsection{Unconditional Audio Generation}\label{sec:unconditional}
Speech and music have rich hierarchies of latent structure.  Speech has complex linguistic structure (phonemes, words, syntax, semantics, etc.) and music has highly compositional musical structure (notes, chords, melody and rhythm, etc.).  The presence of these latent structures in generated samples can be used as a proxy for how well a generative model has learned dependencies at various timescales.  As such, a qualitative analysis of unconditional samples is an insightful method of evaluating generative models of audio.  We train MelNet on three unconditional audio generation tasks---single-speaker speech generation, multi-speaker speech generation, and music generation.  For completeness, the sections below include brief discussions and qualitative observations regarding the generated samples.
However, it is not possible to convey the many characteristics of the generated samples in text and we highly encourage the reader to listen to the audio samples and make their own judgments.  In addition to qualitative analysis, we quantitatively compare MelNet to a WaveNet baseline across each of the three unconditional generation tasks.

\subsubsection{Single-Speaker Speech} To test the model's ability to model a single speaker in a controlled environment, we utilize the Blizzard 2013 dataset~\citep{king2011blizzard}, which consists of audiobook narration performed in a highly animated manner by a professional speaker. We use a 140 hour subset of this dataset for which we were able to find transcriptions, making the dataset also suitable for future text-to-speech experiments.  We find that MelNet frequently generates samples that contain coherent words and phrases.  Even when the model generates incoherent speech, the intonation, prosody, and speaker characteristics remain consistent throughout the duration of the sample.  Furthermore, the model learns to produce speech using a variety of character voices and learns to generate samples which contain elements of narration and dialogue.  Biased samples tend to contain longer strings of comprehensible words but are read in a less expressive fashion.  When primed with a real sequence of audio, MelNet is able to continue sampling speech which has consistent speaking style and intonation.\looseness=-1

\subsubsection{Multi-Speaker Speech} Audiobook data is recorded in a highly controlled environment.  To demonstrate MelNet's capacity to model distributions with significantly more variation, we utilize the VoxCeleb2 dataset~\citep{chung2018voxceleb2}.  The VoxCeleb2 dataset consists of over 2,000 hours of speech data captured with real world noise including laughter, cross-talk, channel effects, music and other sounds.  The dataset is also multilingual, with speech from speakers of 145 different nationalities, covering a wide range  of accents, ages, ethnicities and languages. When trained on the VoxCeleb2 dataset, we find that MelNet is able to generate unconditional samples with significant variation in both speaker characteristics (accent, language, prosody, speaking style) as well as acoustic conditions (background noise and recording quality).  While the generated speech is generally not comprehensible, samples can often be identified as belonging to a specific language, indicating that the model has learned distinct modalities for different languages. Furthermore, it is difficult to distinguish real and fake samples which are spoken in foreign languages.  For foreign languages, semantic structures are not understood by the listener and cannot be used to discriminate between real and fake. Consequently, the listener must rely largely on phonetic structure, which MelNet is able to realistically model.\looseness=-1

\subsubsection{Music} To show that MelNet can model audio modalities other than speech, we apply the model to the task of unconditional music generation.  We utilize the MAESTRO dataset~\citep{hawthorne2018enabling}, which consists of over 172 hours of solo piano performances. The samples demonstrate that MelNet learns musical structures such as melody and harmony.  Furthermore, generated samples often maintain consistent tempo and contain interesting variation in volume, timbre, and rhythm.

\subsubsection{Human Evaluation} Making quantitative comparisons with existing generative models such as WaveNet is difficult for various reasons.  While WaveNet and MelNet both produce exact density estimates, these models cannot be directly compared using log-likelihood as they operate on different representations.  We instead resort to comparing both models by evaluating their ability to model long-range dependencies.  To make this comparison quantitatively, we conduct an experiment where we provide an anonymized ten second sample from both models to human evaluators and ask them to identify the sample which exhibits longer-term structure.  Further details of the methodology for this experiment are provided in Appendix~\ref{sec:human}.  We conduct this experiment for each of the three unconditional audio generation tasks and report results in Table~\ref{table:melwav}.  Evaluators overwhelmingly agreed that samples generated by MelNet had more coherent long-range structure than samples from WaveNet.  Samples from both models are included on the accompanying web page.

In addition to comparing MelNet to an unconditional WaveNet model for music generation, we also compare to a two-stage Wave2Midi2Wave model~\citep{hawthorne2018enabling} which conditions WaveNet on MIDI generated by a separately-trained Music Transformer~\citep{huang2018music}.  Results, shown in Table~\ref{table:melwavmidi}, show that despite having the advantage of directly modelling the musical notes, the two-stage model does not capture long-range structure as well as a MelNet model that is trained entirely end-to-end.

\begin{table}[t]
    \vspace{4pt}

    \centering
    \begin{subtable}[]{.9\linewidth}

        \centering
        \begin{tabular}{l@{\hskip 32pt}c@{\hskip 32pt}c}\toprule
                                & WaveNet   & MelNet   \\\midrule
        Blizzard                & 0.0\hspace{0.12em}\%     & 100.0\hspace{0.12em}\%  \\
        VoxCeleb2               & 0.0\hspace{0.12em}\%     & 100.0\hspace{0.12em}\%  \\
        MAESTRO                 & 4.2\hspace{0.12em}\%     & 95.8\hspace{0.12em}\%   \\\bottomrule
        \end{tabular}
        \vspace{2pt}
        \caption{\label{table:melwav}Comparison between MelNet and WaveNet. Both models are trained in an entirely unsupervised manner.}
    \end{subtable}

    \vspace{16pt}

    \begin{subtable}[]{.9\linewidth}
        \centering
        \begin{tabular}{l@{\hskip 14pt}c@{\hskip 14pt}c}\toprule
                                & Wave2Midi2Wave    & MelNet   \\\midrule
        MAESTRO                   & 37.7\hspace{0.12em}\%            & 62.3\hspace{0.12em}\%   \\\bottomrule
        \end{tabular}
        \vspace{2pt}
        \caption{\label{table:melwavmidi}Comparison between MelNet and Wave2Midi2Wave. Wave2Midi2Wave is a two-stage model consisting of a Music Transformer trained on labelled MIDI followed by a conditional WaveNet model.  The MelNet model, on the other hand, is trained without any intermediate supervision.}
    \end{subtable}

    \vspace{8pt}

    \caption{\label{table:unconditional}Selection rates of human evaluators when asked to identify which model generates samples with longer-term structure.  Results show that MelNet captures long-range structure better than WaveNet.  Furthermore, an end-to-end MelNet model outperforms a two-stage model which conditions WaveNet on generated MIDI.}
    \vspace{1pt}
\end{table}

\subsection{Text-to-Speech Synthesis}
We apply MelNet to the tasks of single-speaker and multi-speaker TTS.  As was done for the unconditional tasks, we provide audio samples on the accompanying web page and provide a brief qualitative analysis of samples in the following sections. We then quantitatively evaluate our text-to-speech models on the task of density estimation.

\subsubsection{Single-Speaker TTS} To assess MelNet's ability to perform the task of single-speaker text-to-speech, we again use the audiobook data from the Blizzard 2013 dataset, including the corresponding transcriptions.  As the dataset contains speech which is spoken in a highly expressive manner with significant variation, the distribution of audio given text is highly multimodal.  To demonstrate that MelNet has learned to model these modalities, we include multiple speech samples for a given text.  The samples demonstrate that MelNet learns to produce diverse vocalizations for the same text, many of which we are unable to easily distinguish from ground truth data.  Furthermore, MelNet learns to infer speaking characteristics from text---samples which contain dialogue are read using various character voices, while narrative text is read in a relatively inexpressive manner.  When primed with a sequence of audio, MelNet effectively infers speaker characteristics and can perform text-to-speech on unseen text while preserving the speaking style of the priming sequence.

\subsubsection{Multi-Speaker TTS} We also train MelNet on a significantly more challenging multi-speaker dataset.  The TED-LIUM 3 dataset~\citep{hernandez2018ted} consists of 452 hours of recorded TED talks. The dataset has various characteristics that make it particularly challenging.  Firstly, the transcriptions are  unpunctuated, unnormalized, and contain errors.  Secondly, speaker IDs are noisy, as they do not discriminate between multiple speakers within a given talk, e.g.\ questions from interviewers and audience members.  Lastly, the dataset includes significant variation in recording conditions, speaker characteristics (over 2,000 unique speakers with diverse accents), and background noise (applause and background music are common).  Despite this, we find that MelNet is capable of producing realistic text-to-speech samples and can generate samples for different speakers by conditioning on different speaker IDs.  Generated samples also contain speech disfluencies (e.g. umms and ahhs), repeated and skipped words, applause, laughter, and various other idiosyncrasies which result from the noisy nature of the data.

\subsubsection{Density Estimation} \label{sec:density}
Generative models trained with maximum-likelihood are most directly evaluated by the likelihood they assign to unseen data. Similar to MelNet, many existing works for end-to-end TTS are designed to model two-dimensional spectral features.  However, density estimates cannot be used straightforwardly for comparison because existing TTS models generally do not use log-likelihood as a training objective.  To make density estimation comparisons possible, we instead define a set of surrogate models which encode assumptions made by existing TTS models and compare the density estimates of these models to our own:\looseness=-1
\begin{itemize}
    \item \textbf{Diagonal Gaussian} The vast majority of end-to-end TTS systems such as Tacotron~\citep{wang2017tacotron}, DeepVoice~\citep{arik2017deep}, VoiceLoop~\citep{taigman2018voiceloop}, Char2Wav~\citep{sotelo2017char2wav}, and ClariNet~\citep{ping2018clarinet} utilize a coarse autoregressive model, where spectral features are factorized as a product of per-frame factors. The elements within each frame are assumed to be conditionally independent and unimodal (given all preceding frames). To represent this class of models, we use a model which factorizes the distribution over frames\looseness=-1
    \begin{equation}
    p(x) = \prod_{i}p(x_{i, *} \mid x_{<i,*})
    \end{equation}
    and models each frame as a diagonal Gaussian with parameterized mean $\mu_i \in \mathbb{R}^{d}$ and standard deviation $\sigma_i \in \mathbb{R}_{+}^{d}$, where $d$ is the dimension of each frame:
    \begin{equation}
    p(x_{i, *} \mid x_{<i, *}) = \mathcal{N}\big(x_{i, *} \pb \mu_i, \text{diag}(\sigma_i^2)\big).
    \end{equation}
    \item \textbf{VAE: Global z} Subsequent works~\citep{akuzawa2018expressive,hsu2018hierarchical} have used a similar frame-level factorization, but utilized a variational autoencoder (VAE)~\citep{kingma2013auto} to jointly model a latent variable $z$ which conditions the generation of each frame. The joint distribution $p(x, z)$ is decomposed as $p(z)p(x \mid z)$, where
    \begin{equation}
    p(x \mid z) = \prod_{i}p(x_{i, *} \mid x_{<i, *}, z).
    \end{equation}
    As before, the conditional distribution over each frame, $p(x_{i, *} \mid x_{<i, *}, z)$, is modelled as a Gaussian with diagonal covariance.
    \item \textbf{VAE: Local z} We also introduce a more expressive VAE which differs only in that it utilizes a sequence of latent variables $z = (z_1, \ldots, z_T)$ instead of a single global latent variable:
    \begin{equation}
    p(x \mid z) = \prod_{i}p(x_{i, *} \mid x_{<i, *}, z_{\leq i}).
    \end{equation}
    \item \textbf{MelNet} To represent the model introduced in this work, we use the probabilistic model described in Section~\ref{sec:model}, as well as a Gaussian variant which simply replaces the GMM with a univariate Gaussian.
\end{itemize}
\begin{table}[t]
\centering
\begin{tabular}{lcc}\toprule
                                            & Unconditional    & Text-to-Speech  \\ \midrule
Diagonal Gaussian                           &  $   = -1.44$    &   $   = -1.56$  \\
VAE: Global $z$                             &  $\leq -1.52$    &   $\leq -1.65$  \\
VAE: Local $z$                              &  $\leq -1.92$    &   $\leq -1.95$  \\
MelNet: Gaussian                            &  $   = -2.29$    &   $   = -2.31$  \\
MelNet: GMM                                 &  $   = -2.32$    &   $   = -2.33$  \\ \bottomrule
\end{tabular}
\caption{\label{table:density} Negative log-likelihood (nats/dim; lower is better) for different probabilistic models on the tasks of unconditional and text-conditional speech generation on the Blizzard dataset.}
\end{table}
We constrain each model to use roughly the same number of parameters and briefly tune hyperparameters to ensure each model is reasonably representative of the potential of each probabilistic model.  We found that the variation resulting from hyperparameters was relatively small in comparison to the margins between different probabilistic models.  Further details for these models can be found in Appendix~\ref{sec:density_deets}.

Results shown in Table~\ref{table:density} demonstrate that fine-grained autoregressive model used by MelNet can greatly improve density estimates for both unconditional speech generation and TTS.  The results also demonstrate that the unimodality and independence assumptions made by existing TTS models are detrimental to density estimates.  Conditioning on latent variables relaxes these independence assumptions and improves performance, though density estimates by VAE models are still inferior to a full autoregressive factorization.  Furthermore, even with a fine-grained factorization, it is beneficial to utilize a multimodal distribution to model the conditional distribution over each element.

\section{Related Work}
The predominant line of research regarding generative models for audio has been directed towards modelling time-domain waveforms with autoregressive models~\citep{oord2016wavenet,mehri2016samplernn,kalchbrenner2018efficient}.
WaveNet is a competitive baseline for audio generation, and as such, is used for comparison in many of our experiments.  However, we note that the contribution of our work is in many ways complementary to that of WaveNet.  MelNet is more proficient at capturing high-level structure, whereas WaveNet is capable of producing higher-fidelity audio.  Several works have demonstrated that time-domain models can be used to invert spectral representations to high-fidelity audio~\citep{shen2018natural,prenger2019waveglow,arik2019fast}, suggesting that MelNet could be used in concert with time-domain models such as WaveNet.

In this work, we tackle the problem of jointly learning global and local structure in an end-to-end manner.  This is in contrast to various works which circumvent the problem of capturing high-level structure by conditioning waveform generation on intermediate features.
Notable such examples include the application of WaveNet to the tasks of text-to-speech~\citep{oord2016wavenet,oord2017parallel,kalchbrenner2018efficient,chen2018sample} and MIDI-conditional music generation~\citep{hawthorne2018enabling,manzelli2018conditioning}.  In the case of TTS, WaveNet depends on a traditional TTS pipeline to produce finely annotated linguistic features (phones, syllables, stress, etc.) as well as pitch and timing information.  In the case of MIDI-conditional music generation, WaveNet relies upon a symbolic music representation (MIDI) which contains the pitch, volume, and timing of notes.
These approaches require datasets with annotated features and are dependent upon human knowledge to determine appropriate domain-specific representations.
In contrast to these approaches, MelNet does not require any intermediate supervision.
MelNet is capable of learning TTS in an entirely end-to-end manner, whereas waveform models have not yet demonstrated the capacity to perform TTS without the assistance of intermediate linguistic features.  Additionally, we demonstrate that MelNet uncovers high-level musical structure as well as two-stage models that separately model intermediate MIDI representations~\citep{hawthorne2018enabling}.

\citet{dieleman2018challenge} and \citet{van2017neural} capture long-range dependencies in waveforms by utilizing a hierarchy of autoencoders.  This approach requires multiple stages of models which must be trained sequentially, whereas the multiscale approach in this work can be parallelized over tiers.   Additionally, these approaches do not directly optimize the data likelihood, nor do they admit tractable marginalization over the latent codes.  We also note that the modelling techniques devised in these works can be broadly applied to autoregressive models such as ours, making their contributions largely complementary to ours.

Recent works have used generative adversarial networks (GANs)~\citep{goodfellow2014generative} to model both waveforms and spectral representations~\citep{donahue2018synthesizing,engel2018gansynth}.
As with image generation, it remains unclear whether GANs capture all modes of the data distribution.
Furthermore, these approaches are restricted to generating fixed-duration segments of audio, which precludes their usage in many audio generation tasks.

Many existing end-to-end TTS models are designed to generate a single high-quality sample for a given text~\citep{arik2017deep,sotelo2017char2wav,wang2017tacotron,ping2018clarinet,taigman2018voiceloop}.
MelNet instead focuses on modelling the full breadth of the conditional distribution of audio given text. We use the task of density estimation to demonstrate that MelNet captures this distribution better than probabilistic models that are commonly used by existing TTS systems, and we show that unimodality and independence assumptions made by existing TTS models are overly restrictive.  Utilizing a more flexible probabilistic model allows MelNet to generate spectrograms with realistic textures without oversmoothing or blurring.  This enables generated spectrograms to be directly inverted to high-fidelity audio using classical spectrogram inversion algorithms, whereas existing spectrogram models which produce audio of comparable quality rely on neural vocoders to correct for oversmoothing~\citep{sotelo2017char2wav,shen2018natural,ping2018clarinet}.

The network architecture used for MelNet is heavily influenced by recent advancements in deep autoregressive models for images. \citet{theis2015generative} introduced an LSTM architecture for autoregressive modelling of 2D images and \citet{oord2016pixel} introduced PixelRNN and PixelCNN and scaled up the models to handle the modelling of natural images.  Subsequent works in autoregressive image modelling have steadily improved state-of-the-art for image density estimation~\citep{van2016conditional,salimans2017pixelcnn++,parmar2018image,chen2017pixelsnail,child2019generating}.  We draw inspiration from many of these models, and ultimately design a recurrent architecture of our own which is suitable for modelling spectrograms rather than images.

We use a multidimensional recurrence in both the time-delayed stack and the upsampling tiers to extract features from two-dimensional inputs.
Our multidimensional recurrence is effectively `factorized' as it independently applies one-dimensional RNNs across each dimension.  This approach differs from the tightly coupled multidimensional recurrences used by MDRNNs~\citep{graves2007multi,graves2009offline} and \mbox{Grid LSTMs~\citep{kalchbrenner2015grid}} and more closely resembles the approach taken by ReNet~\citep{visin2015renet}.  Our approach allows for efficient training as we can extract features from an $M \times N$ grid in $\max(M, N)$ sequential recurrent steps rather than the $M + N$ sequential steps required for tightly coupled recurrences.  Additionally, our approach enables the use of highly optimized one-dimensional RNN implementations.

Various approaches to image generation have succeeded in generating high-resolution, globally coherent images with hundreds of thousands of dimensions~\citep{karras2017progressive,reed2017parallel,kingma2018glow}.  The methods introduced in these works are not directly transferable to waveform generation, as they exploit spatial properties of images which are absent in one-dimensional audio signals.
However, these methods are more straightforwardly applicable to two-dimensional representations such as spectrograms.  Of particular relevance to our work are approaches which combine autoregressive models with multiscale modelling~\citep{oord2016pixel,dahl2017pixel,reed2017parallel,menick2018generating}.  We demonstrate that the benefits of a multiscale autoregressive model extend beyond the task of image generation, and can be used to generate high-resolution, globally coherent spectrograms.

\section{Conclusion}
We have introduced MelNet, a generative model for spectral representations of audio.
MelNet combines a highly expressive autoregressive model with a multiscale modelling scheme to generate high-resolution spectrograms with realistic structure on both local and global scales.
In comparison to previous works which model time-domain signals directly, MelNet is particularly well-suited to model long-range temporal dependencies.
Experiments show promising results on a diverse set of tasks, including unconditional speech generation, music generation, and text-to-speech synthesis.

\section*{Acknowledgements}
We thank Kyle Kastner for reviewing a draft of this paper and providing helpful feedback.

\bibliography{main}
\bibliographystyle{plainnat}

\clearpage
\appendix

\section{Experimental Details}

\subsection{Human Evaluation}\label{sec:human}
For each of the three unconditional audio generation tasks, we generated 50 ten-second samples from WaveNet and 50 ten-second samples from MelNet.  Participants were shown an anonymized, randomly-drawn sample from each model and instructed to ``select the sample which has more coherent long-term structure.''  We collected 50 human evaluations for each task.

\subsection{WaveNet Baseline}
The human evaluation experiments require samples from a baseline WaveNet model.  For the Blizzard and VoxCeleb2 datasets, we use our own reimplementation.  Our WaveNet model uses 8-bit $\mu$-law encoding and models each sample with a discrete distribution. Each model is trained for 150,000 steps.  We use the Adam optimizer~\citep{kingma2014adam} with a learning rate of 0.001 and batch size of 32.  Additional hyperparameters are reported in Table~\ref{table:wavenet}.

\begin{table}[h]
\centering
\begin{tabular}{lccc} \toprule
                     &Blizzard                            &VoxCeleb2   \\ \midrule
Sample Rate (Hz)         & 22,050                          & 16,000 \\
Layers               & 50                              & 60 \\
Kernel Size          & 3                               & 3 \\
Dilation (at layer $i$)            & $2^{i \bmod 10}$                 & $2^{i \bmod 10}$ \\
Receptive Field (samples)     & 10,240                          & 12,288\\
Receptive Field (ms)    & 464                             & 768\\
Max Sample Duration (s)    & 2                             & 2\\
\bottomrule
\end{tabular}
\caption{\label{table:wavenet} WaveNet hyperparameters.}
\end{table}

We do not use our WaveNet implementation for human evaluation on the MAESTRO dataset.  The authors that introduce this dataset provide roughly 2 minutes of audio samples on their website\footnote{\url{https://goo.gl/magenta/maestro-examples}} for both unconditional WaveNet and Wave2Midi2Wave models.  We generate 50 random ten-second slices from these 2 minutes and directly use them for the human evaluations.

\subsection{Density Estimation}\label{sec:density_deets}
We use the Blizzard dataset for all density estimation experiments.  We use low-resolution spectrograms which are typically used by existing TTS systems.  These spectrograms have 80 mel channels and are computed with a STFT hop size of 512 and STFT window size of $6\cdot512$.

All baseline models use similar network architectures which are composed of multi-layer LSTMs with residual connections.  The baseline models use 1024 hidden units whereas the MelNet models use 512 hidden units.  The MelNet models do not use multiscale orderings.  All models have 8-layer autoregressive decoders and the VAE models have an additional 4-layer inference network.  The global VAE model uses a 512-dimensional latent vector and the local VAE model uses a sequence of 32-dimensional latent vectors.  VAE models are trained with KL annealing over the first epoch.  When evaluating density estimates for text-to-speech synthesis, all models use the attention mechanism described in Section~\ref{sec:alignment}.

\end{document}